\documentclass{aa}  

\usepackage{color}
\usepackage{graphicx}
\usepackage{txfonts}
\usepackage{hyperref}

\begin{document} 

\title{Transport coefficients enhanced by suprathermal particles in nonequilibrium heliospheric plasmas}

   \author{E. Husidic
          \inst{1,2}
          \and
          M. Lazar
          \inst{1,2} 
          \and
          H. Fichtner
           \inst{1,3}
          \and K. Scherer
           \inst{1,3} 
           \and
           S. Poedts
           \inst{2,4}
          }

   \institute{Institut für Theoretische Physik, Lehrstuhl IV:
Plasma-Astroteilchenphysik, Ruhr-Universität Bochum, D-44780 Bochum, Germany\\
              \email{eh@tp4.rub.de}
         \and
             Centre for mathematical Plasma Astrophysics/Dept.\ of Mathematics, Celestijnenlaan 200B, 3001 Leuven, Belgium
         \and
             Research Department, Plasmas with Complex Interactions,
Ruhr-Universität Bochum, D-44780 Bochum, Germany
         \and
         Institute of Physics, University of Maria Curie-Sk{\l}odowska, Pl.\ M.\ Curie-Sk{\l}odowska 5, 20-031 Lublin, Poland
             }

   \date{Received: 09 July, 2021; Accepted: 02 August, 2021}

 
  \abstract
   {In heliospheric plasmas, such as the solar wind and planetary magnetospheres, the transport of energy and particles is governed by various fluxes (e.g., heat flux, particle flux, current flow) triggered by different forces, electromagnetic fields, and gradients in density or temperature. In the outer corona and at relatively low heliocentric distances in the solar wind (i.e., < 1~AU), particle-particle collisions play an important role in the transport of energy, momentum, and matter, described within classical transport theory by the transport coefficients, which relate the fluxes to their sources.}
   {The aim of the present paper is to improve the evaluation of the main transport coefficients in such  nonequilibrium plasmas, on the basis of an implicit realistic characterization of their particle velocity distributions, in accord with the in situ observations. Of particular interest is the presence of suprathermal populations and their influence on these transport coefficients.}
   {Using the Boltzmann transport equation and macroscopic laws for the energy and particle fluxes, we derived
   transport coefficients, namely, electric conductivity, thermoelectric coefficient, thermal conductivity, diffusion, and mobility coefficients. These are conditioned by the electrons, which are empirically well described by the Kappa distribution, with a nearly Maxwellian (quasi-thermal) core and power-law tails enhanced by the suprathermal population. Here we have adopted the original Kappa approach that has the ability to outline and quantify the contribution of suprathermal populations.}
{Without exception, the transport coefficients are found to be systematically and markedly enhanced in the presence of suprathermal electrons (i.e., for finite values of the $\kappa$ parameter), due to the additional kinetic energy with which these populations contribute to the dynamics of space plasma systems.  
The present results also show how important an adequate Kappa modeling of suprathermal populations is, which is in contrast to other modified interpretations that underestimate the effects of these populations.}
   {}

   \maketitle
%

\section{Introduction}\label{sec:introduction}
The solar wind is a plasma of a very controversial nature due to the 
multitude of physical processes which act concomitantly or successively 
and control not only local properties, but also the variation of plasma 
parameters with the expansion in heliosphere. 
Indeed, the solar wind is neither a collision-dominated plasma nor a 
collisionless one \citep{Maksimovic-etal-1997, Pierrard-etal-2011}, and 
this is most likely the reason why ideal magnetohydrodynamic (MHD) or collision-free 
(exospheric) models cannot accurately reproduce the global expansion and 
thermodynamics of the solar wind \citep{Maksimovic-etal-1997, 
Matteini-etal-2007, Bale-etal-2013}. Particle-particle collisions are
believed to play an important role at low distances from the Sun in the
outer corona, although there are also wave fluctuations and turbulence
from MHD to proton and electron scales \citep{Bourouaine-etal-2012}. 
At larger heliocentric distances, for example at 1~AU, low-energy particle 
(core) populations remain organized by a so-called collisional age 
\citep{Salem-etal-2003, Stverak-etal-2008, Bale-etal-2009}, while the 
nonequilibrium states of plasma particles with anisotropic velocity (or 
energy) distributions and suprathermal populations 
\citep{Maksimovic-etal-2005, Marsch-2006, Wilson-etal-2019} may be explained by the interaction with wave turbulence and small-scale fluctuations 
\citep{Stverak-etal-2008, Bale-etal-2009, Alexandrova-etal-2013}.
There is observational evidence for the preferential heating
of plasma particles through resonant wave-particle interactions 
\citep{Bourouaine-etal-2011, Wilson-etal-2013}, as well as for the 
generation of more diffuse suprathermal (halo) populations by the wave 
turbulence of a quiet solar wind, or by the instabilities self-induced by the beam (or strahl) populations in the fast solar winds, interplanetary shocks, etc. \citep{Mason-Gloeckler-2012, 
Gurgiolo-etal-2016}. 

Suprathermal populations enhance the high energy tails of the observed 
velocity distributions (up to a few keV) and are characteristic not 
only of the solar wind electrons and protons (see 
\cite{Lazar-etal-2012a} for a review), but they are also reported for 
the heavier (minor) ions at 1~AU  \citep{Collier-etal-1996} and beyond
\citep{Ogilvie-etal-1993}. Theoretical approaches use the so-called
Kappa distributions to describe these populations 
\citep{Pierrard-Lazar-2010} and offer explanations for their origin, 
mainly based on the energization of plasma particles, for example, electrons,
by the interaction with small-scale wave fluctuations, including for instance the electromagnetic 
electron-cyclotron (whistler) waves \citep{Vocks-etal-2008} or the 
electrostatic Langmuir plasma waves \citep{Vinas-etal-2000,Yoon-etal-2006}.
Quasi-linear models involve collisions and a weak plasma 
turbulence, either with predefined spectral profiles 
\citep{Roberts-Miller-1998,Vocks-etal-2008}, or self-supported by 
spontaneous and induced plasma wave fluctuations 
\citep{Yoon-etal-2006}. What matters is that suprathermal populations 
retain the effects of wave turbulence and fluctuations present in the 
system, suggesting that a realistic macroscopic theory for the energy 
and particle transport in space plasmas must rely on the 
assumption that plasma particles are Kappa-distributed. 

Transport processes in a plasma can be studied with a transport equation, such as the (collisional) Boltzmann equation for the evolution of the phase space particle distribution. If a plasma is sufficiently homogeneous, macroscopic equations are obtained by calculating the main velocity moments on the basis of a velocity distribution function describing the plasma particles \citep{Braginskii-1965,Balescu-1988}.  
The macroscopic equations are based on a linear relationship between the fluxes and the external forces and gradients (in, e.g., density, temperature, pressure) driving the fluxes. The constants of proportionality are called transport coefficients, and they contain valuable physical information. The collisional term in the transport equation describes the rate of change of the distribution due to collisions, and by adopting simplified models for the collision term (see Sec.~\ref{sec:transport_coefficients}), one can obtain analytical expressions for the transport coefficients. Because driving gradients and forces lead to deviations from quasi-stationary states, one can assume a small or linear perturbation of the distribution function and linearize the transport equation \citep{Goedbloed-etal-2019}.
For plasmas near thermal equilibrium, the standard Maxwellian distribution is commonly used, while nonthermal particle populations in space plasmas are well described by the family of Kappa distributions (see Sec.~\ref{sec:distributions}).

Recent studies of plasma diffusion, that is, the flux of particles due to a density gradient in the plasma, concerned nonequilibrium plasmas and applied Kappa distributions to determine the diffusion coefficient \citep{Wang-Du-2017,Abbasi-etal-2017}. 
In addition to the diffusion, the mobility coefficient describes the flux of charged particles in the presence of an electrical field and was examined involving Maxwellian \citep{Hagelaar-Pitchford-2005} and Kappa distributions \citep{Wang-Du-2017}. 
The mobility of charged particles is related to the electric conductivity, setting the current density, and the electric field in the relationship \citep{Du-2013, Abbasi-etal-2017}. 
The thermoelectric coefficient relates the electric field to the temperature gradient resulting in electric voltages and currents, and it was derived based on Kappa distributions \citep{Du-2013,Guo-Du-2019}. 
The heat flux is driven by a temperature gradient, for which the thermal conductivity is the constant of proportionality, see, for example, \cite{Rat-etal-2001} for Maxwellian plasma populations, and \cite{Du-2013}, \cite{Guo-Du-2019}, and \cite{Abbasi-Esfandyari-2019} using Kappa distributions. 
However, all of these studies involving Kappa distributions invoke a modified approach that drastically differs from the original Olbertian or the standard Kappa distribution introduced by \cite{Olbert-1968} and \cite{Vasyliunas-1968}. Moreover, recent studies have also pointed to the differences between these two Kappa approaches \citep{Lazar-etal-2015, Lazar-etal-2016}, clearly stating that a  modified Kappa distribution does not allow for a proper comparison with the quasi-thermal core of the observed distributions to outline the suprathermals and their implications (see also Section 2 for more details as well as the discussions in \cite{Lazar-Fichtner-2021}).

Motivated by these precedents, in this paper we propose a re-evaluation of transport coefficients based on a standard Kappa approach, which allows for a correct assessment of the contribution of the suprathermal population. Indeed, here we show that a modified approach systematically leads to an underestimation of the effects of these populations on transport coefficients.
The paper is structured as follows. In Sec.~\ref{sec:distributions} we discuss the applied Kappa power-law distribution function, which describes the suprathermal tails frequently observed in nonthermal space plasmas. Section~\ref{sec:transport_coefficients} presents the calculations for the transport coefficients based on the Kappa distribution, that is to say the electric conductivity, the thermoelectric coefficient, the thermal conductivity, the diffusion coefficient, and the mobility coefficient were derived. We conclude the paper in Sec.~\ref{sec:conclusions} with a summary, discussion, and outlook.

\section{Distributions with suprathermal tails}\label{sec:distributions}

In situ measurements by spacecrafts indicate a ubiquitous presence of suparthermal populations in the velocity (or energy) distributions of plasma particles in the heliosphere \citep{Maksimovic-etal-2005,Zouganelis-etal-2005, Marsch-2006}.
The low-energy core of the measured distributions is nearly Maxwellian (i.e., quasi-thermal), but high-energy tails are suprathermal, decreasing as power laws and markedly departing from the Maxwellian core. Overall the observed distributions are well fitted by the Kappa distribution functions \citep{Pierrard-Lazar-2010}, which are empirically defined by \cite{Olbert-1968} and \cite{Vasyliunas-1968} to describe electrons in the near-Earth solar wind and magnetospheric plasmas.
This model distribution was introduced in the following (isotropic) form 
\begin{equation}\label{eq:skd_general}
    f_\kappa = \frac{n}{\pi^{3/2} \,\Theta^3} 
    \frac{\Gamma(\kappa + 1)}{\kappa^{3/2}\,\Gamma(\kappa - 1/2)}
    \left(1 + \frac{v^2}{\kappa\,\Theta^2} \right)^{-(\kappa + 1)}\,,
\end{equation}
where $n$ is the number density, $v$ is the particle speed, $\kappa$ is the power-law exponent describing the high-energy distribution tails,
$\Gamma (a)$ denotes the (complete) Gamma function of argument $a$, and $\Theta$ was termed the most probable speed, which
relates to the energy of the peak in the differential flux and to the thermal speed of the Maxwellian limit, that is, $\Theta = \Theta_\mathrm{M} \equiv \sqrt{2\,k_\mathrm{B}\,T_\mathrm{M}/m}$ \citep{Olbert-1968, Vasyliunas-1968}, also see below. 
The (kinetic)  temperature for Kappa-distributed populations ($T_\kappa$) is obtained via the second velocity moment
\begin{equation}\label{eq:temperature_integral}
    T_\kappa = \frac{m}{k_\mathrm{B}} \int \mathrm{d}^3v\,v^2\,f_\kappa 
      = \frac{\kappa}{\kappa - 3/2} \frac{m\, \Theta^2}{2\,k_\mathrm{B}}\,\end{equation}
and is defined for $\kappa > 3/2$ (provided that $T_\kappa >0$). Here, $m$ denotes the particle mass and $k_\mathrm{B}$ is Boltzmann's constant. This original form of Kappa distribution allows for a straightforward analysis of suprathermals and their effects by a direct contrast with the (quasi-)thermal core (subscript c) of the observed distribution \citep{Lazar-etal-2015, Lazar-etal-2016, Lazar-Fichtner-2021}. In this case the core is well reproduced by the Maxwellian limit (subscript M)  
\begin{equation}\label{eq:maxwellian}
f_{\rm c} \simeq f_{\rm M} = \lim_{\kappa \to \infty} f_\kappa = \frac{n_{\rm c}}{\pi^{3/2}\,\Theta^3}\,\exp\left(- \frac{v^2}{\Theta^2} \right)
\end{equation}
with the lower temperature 
\begin{equation}
    T_{\rm c} \simeq T_{\rm M}= \frac{m\, \Theta^2}{2\,k_\mathrm{B}} < T_\kappa\,,
\end{equation}
see Eq.~\eqref{eq:temperature_integral}, but with the dominant number density $n_{\rm c} \simeq n = n_{\rm total}$ \citep{Lazar-2017}. The kinetic temperature $T_\kappa$ increases with decreasing $\kappa$, that is, with an enhancing suprathermal population in the tails. Such a contrasting approach has already been applied to outline the effects of suprathermals in various plasma processes, such as spontaneous or induced fluctuations (instabilities), which show a systematic stimulation in the presence of suprathermals \citep{Vinas-etal-2015, Lazar-etal-2015, Lazar-etal-2018, Shaaban-etal-2019, Shaaban-etal-2020}. 

Another modified Kappa distribution, 
\begin{align}\label{eq:skd_simplified}
    f_{\mathrm{m},\kappa} &= \frac{n}{\pi^{3/2}} \left(\frac{m}{2\,k_\mathrm{B}}\right)^{3/2}
    \frac{\Gamma(\kappa + 1)}{T_\kappa^{3/2}\,(\kappa - 3/2)^{3/2}\,\Gamma(\kappa - 1/2)} \nonumber \\
    &\times \left[\frac{m\,v^2/2}{k_\mathrm{B}\,(\kappa - 3/2)\,T_\kappa} \right]^{-(\kappa + 1)}\,,
\end{align}
is also invoked in the literature, including studies on instabilities \citep{Lazar-etal-2011, Lazar-etal-2013, Vinas-etal-2017} and also evaluations of transport coefficients \citep{Du-2013, Guo-Du-2019}. The modified Kappa distribution is expressed in terms of Kappa temperature $T_\kappa$, which is assumed to be a parameter independent of the $\kappa$ exponent 
\footnote{
In this case, parameter $\Theta = \Theta_\mathrm{M}$ differs from $\Theta_{\mathrm{m},\kappa} \equiv \sqrt{(1-3/(2\,\kappa))}\,\Theta$.}. 
This implies that the modified Kappa distribution and its Maxwellian limit obtained for $\kappa \to \infty$ are described by the same temperature, that is, $T_M = T_\kappa$. In that case, the Maxwellian does not reproduce the low-energy core of the modified Kappa distribution.
Therefore, it cannot be used by contrast with Kappa to outline the presence of suprathermals in the tails, and, implicitly, their contribution to various physical properties; see also \cite{Lazar-etal-2015} and \cite{Lazar-etal-2016} for extended comparative analyses. In order to demonstrate how sensitive the evaluations of transport coefficients are to the shape of the distribution, here we adopt a standard (original) Kappa approach \citep{Olbert-1968, Vasyliunas-1968} and compare our results with previous estimations for a Maxwellian-like core, and also for a modified Kappa approach, which underestimates the effects of the suprathermal population.

In order to simplify the computation of the transport coefficients, in Sec.~\ref{sec:transport_coefficients} we assume that the plasma consists of two particle species, ions, and Kappa-distributed electrons. For the derivation of the transport coefficients mainly contributed to by the electrons (e.g., in hot but still collisional plasmas whose electron temperature is comparable to the ion temperature), we rewrite Eq.~\eqref{eq:skd_general} as
\begin{equation}\label{eq:skd_energy}     
f_{\kappa} = N_\mathrm{\kappa} \left(1 + \frac{1}{\eta} \frac{\varepsilon}{k_\mathrm{B}\,T_\mathrm{c}} \right)^{-(\kappa + 1)},
\end{equation}
in terms of the kinetic energy $\varepsilon = m\,v^2/2$, and the normalization constant 
\begin{equation}\label{eq:normalization_constant}
    N_\mathrm{\kappa} = \frac{n}{\pi^{3/2}\,\eta^{3/2}}\,\left(\frac{m}{2\,k_\mathrm{B}\,T_\mathrm{c}}\right)^{3/2} 
    \frac{\Gamma(\kappa + 1)}{\Gamma(\kappa - 1/2)}\,.
\end{equation}
The distinction between the two Kappa approaches is captured by the parameter $\eta$, as for $\eta = \kappa$ the distribution~(\ref{eq:skd_energy}) represents the standard (original) Kappa from~(\ref{eq:skd_general}), and for $\eta = \kappa - 3/2$ it becomes a modified Kappa distribution, as in Eq.~(\ref{eq:skd_simplified}). For the sake of simplicity, in the following we omit the subscript c for the temperature, and adopt $T_\mathrm{c} = T$.

\section{Transport coefficients} \label{sec:transport_coefficients}

Based on these assumptions, we derived the following transport coefficients: the electric conductivity $\sigma$, the thermoelectric coefficient $\alpha$, the thermal conductivity 
$\lambda$, the diffusion coefficient $D$, and the mobility coefficient $\mu$. These parameters can be obtained from the following macroscopic laws:
\begin{align}
    \vec{E} &= \frac{1}{\sigma}\,\vec{j} + 
    \alpha\,\nabla T\,, \label{eq:ohm's_law} \\
    \vec{q} &= (\phi + \alpha\,T)\,\vec{j} 
    - \lambda\, \nabla T\,, \label{eq:fourier's_law} \\
    \vec{j} &= q \int \mathrm{d}^3v\,\vec{v}\,f\,, \label{eq:current_density} \\
    \vec{q} &= \frac{1}{2}m \int \mathrm{d}^3v\,v^2\,\vec{v}\,f
    = \int \mathrm{d}^3v\,\varepsilon\,\vec{v}\,f\,, \label{eq:heat_flux} \\
    \vec{\Gamma} &= \langle \vec{v} \rangle = \int \mathrm{d}^3v\,\vec{v}\,f \,, \label{eq:drift_velocity} \\
    \vec{\Gamma} &= -D\,\nabla n - \mu\,n\,\vec{E}\,, \label{eq:fick's_law} 
\end{align}
where $f$ is the particle distribution, for example the electron Kappa distribution.
Equation~\eqref{eq:ohm's_law} is a generalization of Ohm's law, where $\vec{E}$ denotes 
the electric field and $\vec{j}$ is the current density, which is explicitly defined in 
Eq.~\eqref{eq:current_density}. Equation~\eqref{eq:fourier's_law} is a generalization of Fourier's law, where
$\vec{q}$ is the heat flux, defined in Eq.~\eqref{eq:heat_flux},  
and $\phi$ is the electric potential related to the electric field 
$\vec{E} = -\nabla \phi$. The particle flux defined in (\ref{eq:drift_velocity}) satisfies an extended Fick's law in Eq.~\eqref{eq:fick's_law}, which includes not only the diffusion due to density gradient, but also the mobility of charged particles in the presence of an electric field. For more details, see \cite{Spatschek-1990}, \cite{Boyd-Sanderson-2003}, and \cite{Goedbloed-etal-2019}.

Transport processes in a plasma are in general described starting from the (collisional) Boltzmann equation for the time-space evolution of the distribution function $f$
\begin{equation}\label{eq:boltzmann_eq}
    \frac{\partial f}{\partial t} + \vec{v} \cdot 
    \nabla f + \frac{q}{m} 
    \left[\vec{E} + \frac{1}{c} (\vec{v} \times \vec{B}) \right] 
    \cdot \nabla_{\vec{v}} f = \mathcal{C}(f).
\end{equation}
We note that $\vec{E}$ and $\vec{B}$ are the electric and magnetic  fields, including the imposed and self-generated ones, $c$ is the speed of light in vacuum, $q$ is  the particle's charge, and collisional term $\mathcal{C}(f)$ is in general a nonlinear integral. In the absence of a magnetic field and assuming a stationary transport, Eq.~\eqref{eq:boltzmann_eq} simplifies to
\begin{equation}\label{eq:boltzmann__eq_simplified}
    \vec{v} \cdot \nabla f - \frac{e}{m}\,
    \vec{E} \cdot \nabla_{\vec{v}} f = \mathcal{C}(f)\,.
\end{equation}
The change in the distribution function can be assumed small enough to be described by a linearized velocity distribution
\begin{equation}\label{eq:f_linearized}
    f(\vec{r},\vec{v},t) = f_0(\vec{r},\vec{v}) + f_1(\vec{r},\vec{v},t)\,,
\end{equation}
where $f_0$ is the stationary nonequilibrium distribution and $f_1$ is a small perturbation.
If electrons are hot enough (Lorentzian plasma), we can consider only collisions between the mobile electrons and the stationary, heavy ions. 
For the collisional term, we use a Krook-type collisional operator \citep{Bhatnagar-etal-1954} and
assume that the disturbed distribution relaxes under the effects of collisions with a rate 
$\nu_\mathrm{ei}(v)$, which allows us to write 
\begin{equation}\label{eq:krook_collision_term}
    \mathcal{C}(f) = - \nu_\mathrm{ei}(v)\,(f - f_0) = - \nu_\mathrm{ei}(v)\,f_1\,.
\end{equation}
The collision frequency $\nu_\mathrm{ei}(v)$ between electrons and ions (subscript "ei") can be described by
\citep{Helander-Sigmar-2005}
\begin{equation}\label{eq:collision_frequency}
    \nu_\mathrm{ei}(v) = \nu_\mathrm{ei} = 
    \frac{4\,\pi\,n_\mathrm{e}\,z\,e^4\,L_\mathrm{ei}}{m^2\,v^3}\,,
\end{equation}
where $z$ denotes the ion charge number, $L_\mathrm{ei}$ is the Coulomb logarithm 
$L_\mathrm{ei} \equiv \ln{\Lambda}$, and $\Lambda$ is the electron Debye length (normalized
to the impact parameter). 
Inserting Eq.~\eqref{eq:krook_collision_term} into Eq.~\eqref{eq:boltzmann__eq_simplified} 
and setting $f_0 = f_\kappa$ yields
\begin{equation}\label{eq:boltzmann_with_coll_freq}
     \vec{v} \cdot \nabla f_\kappa - \frac{e}{m}\,
    \vec{E} \cdot \nabla_{\vec{v}} f_\kappa = -\nu_\mathrm{ei}\,f_1\,,
\end{equation}
where the spatial and velocity gradients of the perturbation of second order can be neglected. By using Eq.~\eqref{eq:skd_energy}, we can transform Eq.~\eqref{eq:boltzmann_with_coll_freq}
into
\begin{equation}\label{eq:f_1}
  \begin{split}
    f_1 =& 
    - \frac{f_\kappa\,(\kappa + 1)}{\nu_\mathrm{ei}\,\left(k_\mathrm{B}\,T\,\eta + \varepsilon\right)}
    \left(e\,\vec{E} + \frac{\varepsilon}{T}\, \nabla T\right) \cdot \vec{v} \\
    &+ \frac{3}{2} \frac{f_\kappa}{\nu_\mathrm{ei}} \frac{\nabla T}{T} \cdot \vec{v} 
    - \frac{f_\kappa}{\nu_\mathrm{ei}\,n} \nabla n \cdot \vec{v}\,.
  \end{split}    
\end{equation}
We note that the middle (energy-free) term with the temperature gradient does not appear in Eq.~(10) from \cite{Du-2013}. This term does not affect the computation of the electric conductivity $\sigma$, the diffusion coefficient $D$, or the mobility coefficient $\mu$, but it can have an important influence on the thermoelectric coefficient $\alpha$ and thermal conductivity $\lambda$, see the Appendix \ref{app:correction}.

\subsection{The electric conductivity} \label{subsec:sigma}

The electric conductivity $\sigma$ was calculated by setting $\nabla T = \vec{0}$ and $\nabla n = \vec{0}$, reducing the perturbed distribution in Eq.~\eqref{eq:f_1} to
\begin{equation}\label{eq:f_1_for_sigma}
    f_1 = 
    - \frac{f_\kappa\,(\kappa + 1)\,e}{\nu_\mathrm{ei}\,\left(k_\mathrm{B}\,T\,\eta + \varepsilon\right)}
    \vec{E} \cdot \vec{v}\,.
\end{equation}
By inserting \eqref{eq:f_linearized} into \eqref{eq:current_density}, we obtain
\begin{equation}\label{eq:current_density_with_f_1_for_sigma}
    \vec{j} = -e \int \mathrm{d}^3v\,\vec{v}\,f
    = -e \int \mathrm{d}^3v\,\vec{v}\,f_1\,
\end{equation}
(using $\int \mathrm{d}^3v\,\vec{v}\,f_\kappa = 0$, due to the odd integrand as a function of velocity).
In Eq.~\eqref{eq:current_density_with_f_1_for_sigma}, we replace $f_1$ with the expression in Eq.~\eqref{eq:f_1_for_sigma} and find
\begin{equation}\label{eq:current_density_with_average_for_sigma}
  \begin{split}
    \vec{j} &= (\kappa + 1)\,e^2\,\int \mathrm{d}^3v\,
    \frac{\vec{v} \vec{v}}{\nu_\mathrm{ei}} 
    \left(k_\mathrm{B}\,T\,\eta + \varepsilon \right)^{-1} f_\kappa \,\vec{E}  \\
    &= \frac{(\kappa + 1)\,e^2}{3}\,
    \Bigg\langle \frac{v^2}{\nu_\mathrm{ei}} 
    \left(k_\mathrm{B}\,T\,\eta + \varepsilon \right)^{-1} \Bigg\rangle \,\vec{E}\,.
  \end{split}
\end{equation}
Here, $\vec{v} \vec{v}$ denotes the dyadic product, $\mathcal{I}$ is the unit tensor, and the integral can be written in the form $\int \mathrm{d}^3v\,
\vec{v} \vec{v}\,G(v) = 1/3\,\mathcal{I} \int \mathrm{d}^3v\,v^2\,G(v)$ as the cross diagonal terms vanish for any $G(v)$ as a function of speed $v$ given by isotropic distribution. This was also applied in the next derivations. We further introduced the average $\langle F(v) \rangle \equiv \int \mathrm{d}^3v\,F(v)\,f_\kappa$, where $F(v)$ is some function of $v$. 
Comparing Eqs.~\eqref{eq:current_density_with_average_for_sigma} and \eqref{eq:ohm's_law} yields the electric conductivity
\begin{equation}\label{eq:sigma_with_average}
    \sigma = \frac{(\kappa + 1)\,e^2}{3}
    \Bigg\langle \frac{v^2}{\nu_\mathrm{ei}} 
    \left(k_\mathrm{B}\,T\,\eta + \varepsilon \right)^{-1} \Bigg\rangle\,,
\end{equation}
and after inserting the expressions for the kinetic energy $\varepsilon$ as well as the collision frequency from Eq.~\eqref{eq:collision_frequency}, 
this reads
as follows:%
\begin{equation}\label{eq:current_density_with_coll_freq}
    \sigma = \frac{(\kappa + 1)\,m^2}{12\,\pi\,k_\mathrm{B}\,T\,n\,z\,e^2\,L_\mathrm{ei}\,\eta}
    \Big \langle v^5\, (1 + A_\kappa\,v^2)^{-1} \Big \rangle
,\end{equation}
with $A_\kappa \equiv m/(2\,k_\mathrm{B}\,T\,\eta)$. By solving the average value (see 
Appendix~\ref{app:Integrals}), the final expression for $\sigma$ becomes
\begin{equation}\label{eq:sigma}
   \sigma = 
   \frac{\Gamma\left(\kappa - 2\right)}{\Gamma\left(\kappa - 1/2\right)}\, \eta^{3/2}\,
   \frac{4\,\sqrt{2}\,\left(k_\mathrm{B}\,T \right)^{3/2}}{m^{1/2}\,\pi^{3/2}\,z\,e^2\,L_\mathrm{ei}}
\end{equation}
with $\eta = \kappa$ for the standard Kappa approach and $\eta = \kappa - 3/2$ for the modified one. For 
$\kappa \to \infty$, the Maxwellian limit is recovered:
\begin{equation}
    \sigma_\mathrm{M} = \frac{4\,\sqrt{2}\,
    \left(k_\mathrm{B}\,T \right)^{3/2}}{m^{1/2}\,\pi^{3/2}\,z\,e^2\,L_\mathrm{ei}}.
\end{equation}

\begin{figure}
   \includegraphics[width=\hsize]{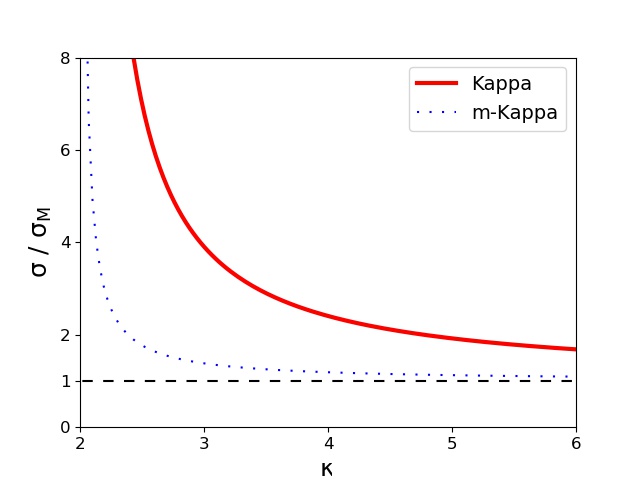}
   \caption{Electric conductivity $\sigma$ as a function of $\kappa$, for the standard Kappa (red) and the modified Kappa (m-Kappa, blue-dotted) approaches, and normalized to the Maxwellian limit $\sigma_\mathrm{M}$ (which is reduced to the horizontal dashed line). The two Kappa-based functions approach the Maxwellian limit for $\kappa \to \infty$ and diverge at the pole $\kappa = 2$.}
              \label{fig:sigma}%
    \end{figure}

Figure~\ref{fig:sigma} displays the normalized electric conductivity as a function of $\kappa$ for both Kappa approaches, 
the standard one (red line) and the modified one (m-Kappa, with a blue-dotted line), and also for the Maxwellian limit (black-dashed line). In order to outline the differences
between these three approaches, we normalized all of the results to the Maxwellian limit. With the increase in the $\kappa$ parameter (i.e., $\kappa \to
\infty$), both Kappa approaches lead to the Maxwellian limit. However, the modified Kappa decreases much faster and does not show much
of a difference from this limit, except for only very low values of $\kappa$ approaching the singularity (or pole) $\kappa =2$.
However, the standard Kappa approach shows significant contributions from the suprathermals electrons, which markedly
enhance the electric conductivity even for moderate values $\kappa < 5$; readers are encouraged to compare the red- and black-dashed lines and the corresponding results
in Tables~\ref{tab:c1_sigma_mu} in Appendix~\ref{app:values}. The underestimations obtained with a modified Kappa approach are also
significant, as shown by the comparison of red- and blue-dotted lines, and the corresponding values in Table~\ref{tab:c1_sigma_mu} in Appendix~\ref{app:values}.
%

\subsection{The thermoelectric coefficient} \label{subsec:alpha}
The thermoelectric coefficient $\alpha$ was derived by setting $\vec{E} = \vec{0}$ and $\nabla n = \vec{0}$, which simplifies Eq.~\eqref{eq:f_1} to
\begin{equation}\label{eq:f_1_for_alpha}
    f_1 = 
    - \frac{f_\kappa\,(\kappa + 1)}{\nu_\mathrm{ei}\,\left(k_\mathrm{B}\,T\,\eta + \varepsilon\right)}
    \frac{\varepsilon}{T}\, \nabla T \cdot \vec{v}
    + \frac{3}{2} \frac{f_\kappa}{\nu_\mathrm{ei}} \frac{\nabla T}{T} \cdot \vec{v}\,.
\end{equation}
We inserted Eq.~\eqref{eq:f_1_for_alpha} into \eqref{eq:current_density} and obtain
\begin{equation}\label{eq:current_density_for_alpha}
  \begin{split}
    \vec{j} = & \frac{(\kappa + 1)\,e}{T} \nabla T \cdot 
    \int \mathrm{d}^3v\,
    \frac{\vec{v}\vec{v}}{\nu_\mathrm{ei}} 
    \left(k_\mathrm{B}\,T\,\eta + \varepsilon \right)^{-1} \varepsilon\, f_\kappa \\
    &- \frac{3}{2} \frac{e}{T} \nabla T \cdot \int \mathrm{d}^3v\,
    \frac{\vec{v} \vec{v}}{\nu_\mathrm{ei}} f_\kappa \\
    = & \left[\frac{(\kappa + 1)\,e}{3\,T} 
    \Bigg\langle \frac{v^2\,\varepsilon}{\nu_\mathrm{ei}} 
    \left(k_\mathrm{B}\,T\,\eta + \varepsilon \right)^{-1} \Bigg\rangle
    - \frac{e}{2\,T} \Bigg\langle \frac{v^2}{\nu_\mathrm{ei}} \Bigg\rangle \right] \nabla T\,.
  \end{split}
\end{equation}
From Eqs.~\eqref{eq:current_density_for_alpha} and \eqref{eq:ohm's_law}, we can identify the expression for $\alpha$ as
\begin{equation}\label{eq:alpha_with_average}
  \begin{split}
    \alpha = & - \frac{(\kappa + 1)\,e}{3\,\sigma\,T}
     \Bigg\langle \frac{v^2\,\varepsilon}{\nu_\mathrm{ei}} 
    \left(k_\mathrm{B}\,T\,\eta + \varepsilon \right)^{-1} \Bigg\rangle
    + \frac{e}{2\,\sigma\,T} \Bigg\langle \frac{v^2}{\nu_\mathrm{ei}} \Bigg\rangle \\
    = &  -\frac{(\kappa + 1)\,m^3}{24\,\pi\,\sigma\,k_\mathrm{B}\,T^2\,\,n\,\eta\,z\,e^3\,L_\mathrm{ei}}
    \Big \langle v^7\, (1 + A_\kappa\,v^2)^{-1} \Big \rangle \\
    &+ \frac{m^2}{8\,\pi\,\sigma\,T\,n\,z\,e^3\,L_\mathrm{ei}} \Big \langle v^5 \Big \rangle\,.
\end{split}
\end{equation}
Solving the integrals (see Appendix~\ref{app:Integrals}) and inserting the expression for $\sigma$ from Eq.~\eqref{eq:sigma} yields the final form of the thermoelectric coefficient
\begin{equation}\label{eq:alpha}
        \alpha = -\frac{\eta}{\kappa - 3} \frac{5}{2} \frac{k_\mathrm{B}}{e}\,,
\end{equation}
with $\eta = \kappa$ for the standard Kappa. For the modified Kappa approach ($\eta =
\kappa - 3/2$), the result in Eq.~\eqref{eq:alpha} slightly differs from Eq.~(22) in \cite{Du-2013}, and this is due to the correction term in
Eq.~\eqref{eq:f_1}. A graphical comparison of Eq.~\eqref{eq:alpha} with the result in \cite{Du-2013} is shown in 
Appendix~\ref{app:correction}. The Maxwellian limit ($\kappa \to \infty$) takes the following form:
\begin{equation}
    \alpha_\mathrm{M} = -\frac{5}{2} \frac{k_\mathrm{B}}{e}\,.
\end{equation}

\begin{figure}
   \includegraphics[width=\hsize]{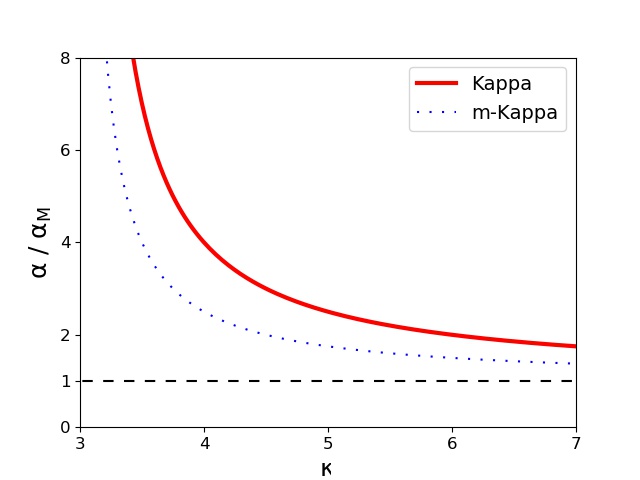}
   \caption{Thermoelectric coefficient $\alpha$ as a function of $\kappa$, for the standard Kappa (red) and the modified Kappa (m-Kappa, blue-dotted) approaches, and normalized to the Maxwellian limit $\alpha_\mathrm{M}$ (which is reduced to the horizontal dashed line). The two Kappa-based functions approach the Maxwellian limit for $\kappa \to \infty$ and diverge at the pole $\kappa = 3$.}
              \label{fig:alpha}%
    \end{figure}

In Fig.~\ref{fig:alpha} the normalized thermoelectric coefficient is plotted as a function of the $\kappa$ parameter for both  
Kappa approaches, the standard one (red line) and modified one (blue-dotted line), and these are compared to the Maxwellian limit (black-dashed line). Again,
for the normalization, we used the same Maxwellian limit. For large $\kappa \to \infty$, both Kappa approaches tend to be similar to the Maxwellian result, while for
finite values of $\kappa < 6$ (but greater than pole at $\kappa = 3$), the standard Kappa approach ($\eta = \kappa$) yields markedly larger values of $\alpha$ (readers are encouraged to compare the red- and black-dashed lines and 
selected values in Table~\ref{tab:c2_alpha} in Appendix~\ref{app:values}). Instead, a modified Kappa underestimates the effects of
suprathermals on $\alpha$, as has resulted from a comparison of red- and blue-dotted lines, and the corresponding results in Table~\ref{tab:c2_alpha} in Appendix~\ref{app:values}.

%

\subsection{The thermal conductivity} \label{subsec:lambda}
For the evaluation of thermal conductivity ($\lambda$) in the absence of an electric current ($\vec{j} = \vec{0}$), Eq.~\eqref{eq:ohm's_law} becomes $\vec{E} = \alpha\,\nabla T$, which we inserted into Eq.~\eqref{eq:f_1} (and set $\nabla n = \vec{0}$) to obtain
\begin{equation}\label{eq:f_1_for_lambda}
    f_1 = 
    - \frac{f_\kappa\,(\kappa + 1)\,(\alpha\,e + \varepsilon/T)}
    {\nu_\mathrm{ei}\,\left(k_\mathrm{B}\,T\,\eta + \varepsilon\right)}
    \, \nabla T \cdot \vec{v}
    + \frac{3}{2} \frac{f_\kappa}{\nu_\mathrm{ei}} \frac{\nabla T}{T} \cdot \vec{v}\,.
\end{equation}
In Eq.~\eqref{eq:heat_flux}, for the heat flux, the integral of the stationary $f_\kappa$ vanishes (integrand is an odd function of $v$), and by substituting $f_1$  from Eq.~\eqref{eq:f_1_for_lambda}, the heat flux becomes
\begin{align}\label{eq:heat_flux_for_lambda}
        \vec{q} =& -(\kappa + 1)\, \int \mathrm{d}^3v\,
    \frac{\vec{v} \vec{v}}{\nu_\mathrm{ei}} 
    \frac{(\alpha\,e + \varepsilon/T)\,\varepsilon\,f_\kappa}{k_\mathrm{B}\,T\,\eta + \varepsilon}\,\nabla T \nonumber \\
    &+ \frac{3}{2\,T} \int \mathrm{d}^3v\,
    \frac{\vec{v} \vec{v}}{\nu_\mathrm{ei}} \varepsilon\,f_\kappa\,\nabla T \nonumber \\
    = & \left(- \frac{\kappa + 1}{3} \Bigg\langle \frac{v^2}{\nu_\mathrm{ei}} 
    \frac{(\alpha\,e + \varepsilon/T)\,\varepsilon}{k_\mathrm{B}\,T\,\eta + \varepsilon} \Bigg\rangle
    + \frac{1}{2\,T} \Bigg\langle \frac{v^2}{\nu_\mathrm{ei}} \varepsilon \Bigg\rangle \right) \nabla T\,.
\end{align}
Comparing the coefficients of $\nabla T$ in Eqs.~\eqref{eq:heat_flux_for_lambda} and 
\eqref{eq:fourier's_law} allowed us to identify the thermal conductivity
\begin{align}\label{eq:lambda_with_average}
    \lambda = & \frac{\kappa + 1}{3} \Bigg\langle \frac{v^2}{\nu_\mathrm{ei}} 
    \frac{(\alpha\,e + \varepsilon/T)\,\varepsilon}{k_\mathrm{B}\,T\,\eta + \varepsilon} \Bigg\rangle
    - \frac{1}{2\,T} \Bigg\langle \frac{v^2}{\nu_\mathrm{ei}} \varepsilon \Bigg\rangle \nonumber \\
     = & \frac{(\kappa + 1)\,\alpha\,m^3}{24\,\pi\,z\,e^3\,k_\mathrm{B}\,T\,n\,\eta\,L_\mathrm{ei}}
    \Big \langle v^7\,(1 + A_\kappa\,v^2)^{-1} \Big \rangle \nonumber \\
    &+ \frac{(\kappa + 1)\,m^4}{48\,\pi\,z\,e^4\,k_\mathrm{B}\,T^2\,n\,\eta\,L_\mathrm{ei}}
    \Big \langle v^9\,(1 + A_\kappa\,v^2)^{-1} \Big \rangle \nonumber \\
    &- \frac{m^3}{16\,\pi\,z\,e^4\,n\,T\,L_\mathrm{ei}} \Big \langle v^7 \Big \rangle \,,
\end{align}
where the collision frequency from Eq.~\eqref{eq:collision_frequency} was 
inserted. After solving the integrals in Eq.~\eqref{eq:lambda_with_average} (see Appendix~\ref{app:Integrals}), we obtain the final expression for the thermal conductivity
\begin{equation}\label{eq:lambda}
    \lambda = \frac{\Gamma(\kappa - 4)}{\Gamma\left(\kappa - 1/2\right)}\,
    \frac{\eta^{7/2}}{(\kappa - 3)}\,\left(16\,\kappa - 8 \right)\,
    \frac{\sqrt{2}\,k_\mathrm{B}\,\left(k_\mathrm{B}\,T \right)^{5/2}}
    {m^{1/2}\,\pi^{3/2}\,z\,e^4\,L_\mathrm{ei}}\,,
\end{equation}
with $\eta = \kappa$ for the standard Kappa approach. For the modified Kappa ($\eta = \kappa - 3/2$), the expression obtained for $\lambda$  differs from Eq.~(26) in \cite{Du-2013} 
due to the additional middle term in Eq.~\eqref{eq:f_1}. A 
graphical contrast from Eq.~\eqref{eq:lambda} to the result in \cite{Du-2013} is displayed in Appendix~\ref{app:correction}.
In this case, the Maxwellian limit ($\kappa \to \infty$) takes the following form:
\begin{equation}
    \lambda_\mathrm{M} = \frac{16\,\sqrt{2}\,k_\mathrm{B}\,\left(k_\mathrm{B}\,T \right)^{5/2}}
    {m^{1/2}\,\pi^{3/2}\,z\,e^4\,L_\mathrm{ei}}\,.
\end{equation}
   \begin{figure}[t!]
   \includegraphics[width=\hsize]{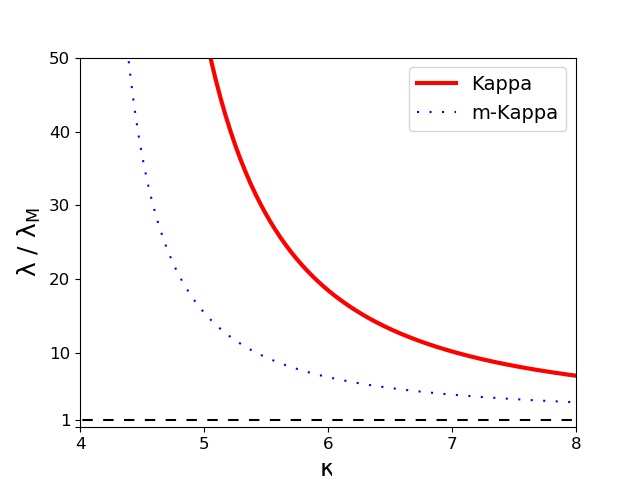}
   \caption{Thermal conductivity $\lambda$ as a function of $\kappa$, for the standard Kappa (red) and the modified Kappa (m-Kappa, blue-dotted) approaches, and normalized to the Maxwellian limit $\lambda_\mathrm{M}$ (which is reduced to the horizontal dashed line). The two Kappa-based functions approach the Maxwellian limit for $\kappa \to \infty$ and diverge at the pole $\kappa = 4$.} \label{fig:lambda}%
    \end{figure}

Thermal conductivity is plotted in Fig.~\ref{fig:lambda} as a function of $\kappa$, for both Kappa approaches, and normalized to the Maxwellian limit.
While for very large values of $\kappa \to \infty$, both Kappa approaches approach their Maxwellian limit, for finite and still large $\kappa < 8$ (but greater than the pole at $\kappa = 4$),
the standard Kappa provides estimations of $\lambda$ (red line) much greater than those
obtained with a modified Kappa approach (blue-dotted line). By contrasting this to the Maxwellian limit (black-dashed line), we also find a
significant contribution of suprathermal population to thermal conductivity. See also the corresponding results in Table~\ref{tab:c3_lambda} in
Appendix~\ref{app:values}.

\subsection{The diffusion coefficient} \label{subsec:diffusion}
The diffusion coefficient $D$ was obtained from Eq.~\eqref{eq:f_1} by setting both $\vec{E} = \vec{0}$ and $\nabla T = \vec{0}$, such that
\begin{equation}\label{eq:f_1_for_diffusion}
    f_1 = - \frac{f_\kappa}{\nu_\mathrm{ei}\,n} \nabla n \cdot \vec{v}\,.
\end{equation}
The particle flux from Eq.~\eqref{eq:drift_velocity} simplifies to
\begin{equation}\label{eq:diffusion_integral_general}
    \vec{\Gamma} = \int \mathrm{d}^3v\,\vec{v}\,f 
    = \int \mathrm{d}^3v\,\vec{v}\,f_1\,,
\end{equation}
as the integral $\int \mathrm{d}^3v\,\vec{v}\,f_\kappa$ vanishes.
We inserted Eq.~\eqref{eq:f_1_for_diffusion} into \eqref{eq:diffusion_integral_general} to obtain
\begin{equation}\label{eq:drift_velocity_with_average_for_diffusion}
    \vec{\Gamma}= - \frac{1}{n}\, \int \mathrm{d}^3v\,
    \frac{\vec{v} \vec{v}}{\nu_\mathrm{ei}} f_\kappa \nabla n = - \frac{1}{3\,n}\,
    \Bigg\langle \frac{v^2}{\nu_\mathrm{ei}} \Bigg\rangle \,\nabla n\,,
\end{equation}
and comparing the coefficients of $\nabla n$ in Eqs.~\eqref{eq:drift_velocity_with_average_for_diffusion} and \eqref{eq:fick's_law} allowed us to identify the diffusion coefficient
\begin{equation}\label{eq:D_with_average}
    D = \frac{1}{3\,n}\,
    \Bigg\langle \frac{v^2}{\nu_\mathrm{ei}} \Bigg\rangle\
    = \frac{m^2}{12\,\pi\,z\,e^4\,L_\mathrm{ei}\,n^2}\, \Big \langle v^5 \Big \rangle\,.
\end{equation}
The computation of the integral in Eq.~\eqref{eq:D_with_average} (see Appendix~\ref{app:Integrals}) yields the final form of the diffusion coefficient
\begin{equation}\label{eq:diffusion_coefficient}
    D = \frac{\Gamma(\kappa - 3)}{\Gamma(\kappa - 1/2)}\,\eta^{5/2}\,
    \frac{4\,\sqrt{2}\,(k_\mathrm{B}\,T)^{5/2}}{\pi^{3/2}\,m^{1/2}\,z\,e^4\,L_\mathrm{ei}\,n}\,,
\end{equation}
with a singularity at $\kappa = 3$, and with $\eta = \kappa$ for the standard Kappa and $\eta = \kappa - 3/2$ for the  modified one. For $\kappa \to \infty$, Eq.~\eqref{eq:diffusion_coefficient} leads to the Maxwellian limit
\begin{equation}\label{eq:diffusion_maxwellian}
    D_\mathrm{M} = \frac{4\,\sqrt{2}\,(k_\mathrm{B}\,T)^{5/2}}{\pi^{3/2}\,m^{1/2}\,z\,e^4\,L_\mathrm{ei}\,n}\,.
\end{equation}
   \begin{figure}
   \includegraphics[width=\hsize]{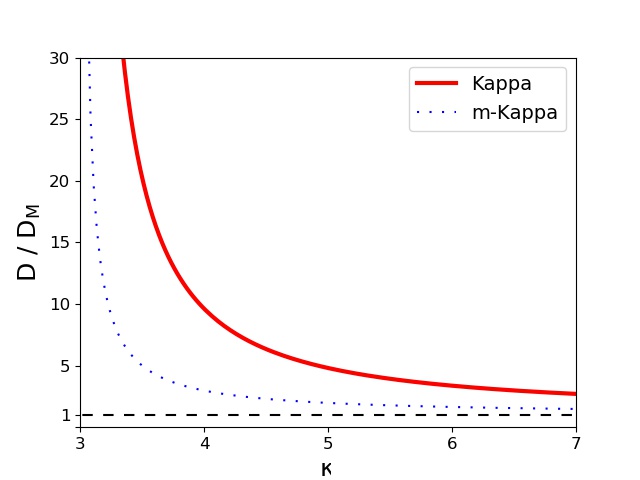}
   \caption{Diffusion coefficient $D$ as a function of $\kappa$, for the standard Kappa (red) and the modified Kappa (m-Kappa, blue-dotted) approaches, and normalized to the Maxwellian limit $\alpha_\mathrm{M}$ (which is reduced to the horizontal dashed line). The two Kappa-based functions approach the Maxwellian limit for $\kappa \to \infty$ and diverge at the pole $\kappa = 3$.}
              \label{fig:diffusion}%
    \end{figure}

Figure~\ref{fig:diffusion} displays the diffusion coefficient as a function of $\kappa$ for both Kappa approaches, the standard (red line) and modified one (blue-dotted line). These are normalized to the Maxwellian limit ($\kappa \to \infty$), which reduces to the horizontal dashed line. Similar to the previous results, we find the results from a modified Kappa approach  ($\eta = \kappa - 3/2$) to decrease faster to the Maxwellian limit and to underestimate the diffusion coefficient  overall, by comparison to the standard Kappa approach ($\eta = \kappa$). Values tabulated for the diffusion coefficient in Table \ref{tab:c4_diffusion} in Appendix~\ref{app:values} may also help to understand these differences.

\subsection{The mobility coefficient} \label{subsec:mobility}
The mobility coefficient $\mu$ was derived by setting $\nabla T = \vec{0}$ and $\nabla n = \vec{0}$ 
in Eq.~\eqref{eq:f_1}, resulting in the expression for the electric conductivity $\sigma$ in Eq.~\eqref{eq:f_1_for_sigma}. 
Thus, we inserted Eq.~\eqref{eq:f_1_for_sigma} into Eq.~\eqref{eq:diffusion_integral_general} to obtain
\begin{equation}\label{eq:drift_velocity_with_average_for_mu}
    \vec{\Gamma} = - \frac{(\kappa + 1)\,e}{3}\,
    \Bigg\langle \frac{v^2}{\nu_\mathrm{ei}} 
    \left(k_\mathrm{B}\,T\,\eta + \varepsilon \right)^{-1} \Bigg\rangle \,\vec{E}\,.
\end{equation}
This expression is similar to 
Eq.~\eqref{eq:current_density_with_average_for_sigma}, containing the same integral. We can, therefore, write the mobility coefficient as
\begin{equation}\label{eq:mobility_coefficient}
    \mu = \frac{\Gamma\left(\kappa - 2\right)}{\Gamma\left(\kappa - 1/2\right)}\, \eta^{3/2}
    \frac{4\,\sqrt{2}\,(k_\mathrm{B}\,T)^{3/2}}{\pi^{3/2}\,m^{1/2}\,z\,e^3\,L_\mathrm{ei}\,n}\,,
\end{equation}
which has the same pole, that is, $\kappa = 2$, as the electric conductivity $\sigma$. Again, $\eta = \kappa$ yields the Olbertian, while $\eta = \kappa - 3/2$ yields the modified approach.  For $\kappa \to \infty$, we find the Maxwellian limit
\begin{equation}\label{eq:mobility_maxwellian}
    \mu_\mathrm{M} = \frac{4\,\sqrt{2}\,(k_\mathrm{B}\,T)^{3/2}}{\pi^{3/2}\,m^{1/2}\,z\,e^3\,L_\mathrm{ei}\,n}\,.
\end{equation}

\begin{figure} [t!]
   \includegraphics[width=\hsize]{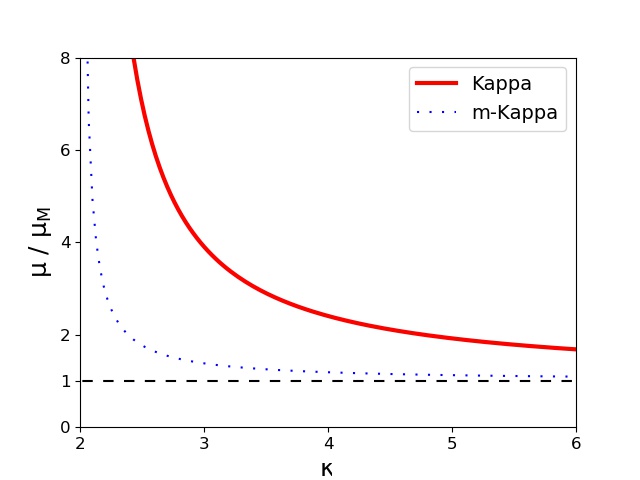}
   \caption{Mobility coefficient $\mu$ as a function of $\kappa$, for the standard Kappa (red) and the modified Kappa (m-Kappa, blue-dotted) approaches, and normalized to the Maxwellian limit $\alpha_\mathrm{M}$ (which is reduced to the horizontal dashed line). The two Kappa-based functions approach the Maxwellian limit for $\kappa \to \infty$ and diverge at the pole $\kappa = 2$.}
              \label{fig:mobility}%
    \end{figure}

Figure~\ref{fig:mobility} shows the mobility coefficient as a function of $\kappa$ for both Kappa approaches, the standard (red line) and the modified one (blue-dotted line), normalized to the Maxwellian limit. The enhancement of $\mu$ in the presence of suprathermals (i.e., in the standard Kappa), but also the underestimates from a modified Kappa, are significant, as previously discussed  in Fig.~\ref{fig:sigma} for the electric conductivity $\sigma$. Selected values for $\mu$ are tabulated in Table \ref{tab:c5_mu} in Appendix~\ref{app:values}, and they make these differences obvious.

From Eqs.~\eqref{eq:sigma} and \eqref{eq:mobility_coefficient}, we obtain the familiar relation between
the electric conductivity and the mobility coefficient
\begin{equation}\label{eq:relation_mu_sigma}
    \sigma = n\,e\,\mu\,.
\end{equation}
Furthermore, when combining Eqs.~\eqref{eq:diffusion_coefficient} and \eqref{eq:mobility_coefficient}, we find the Einstein relation for charged particles
\begin{equation}\label{eq:einstein_relation}
    D = \frac{\eta}{\kappa - 3}\,\frac{ \mu\,k_\mathrm{B}\,T}{e}\,.
\end{equation}
For $\kappa \to \infty$ this reduces to the familiar Einstein relation for charged particles for  Maxwellian populations, that is, $D_\mathrm{M} = \mu_\mathrm{M}\,k_\mathrm{B}\,T/e$.

%

\section{Conclusions and outlook} \label{sec:conclusions}
In the present paper we calculated transport coefficients in nonequilibrium plasmas and have 
shown the consequences of the presence of suprathermal particles on these coefficients. 
By assuming a plasma with Kappa-distributed electrons and using a particular 
expression for the collision frequency, we derived expressions for the electric conductivity 
$\sigma$, the thermoelectric coefficient $\alpha$, the thermal conductivity $\lambda$, the 
diffusion coefficient $D$ and the mobility coefficient $\mu$. The results indicate  
significant enhancements in the transport coefficients in the presence of suprathermal particles.
For the Kappa distribution,  two common approaches are used in literature, the standard Kappa distribution  
originally introduced by \cite{Olbert-1968} and \cite{Vasyliunas-1968}, which enables a proper description of suprathermal populations in contrast to the quasi-thermal core, and a modified Kappa distribution, which does not have the same relevance (see section 2). We computed the results for both Kappa approaches and have shown that the modified 
Kappa approach consistently underestimates the transport coefficients by comparison to the 
original Kappa approach at small values of $\kappa$, while the transport coefficients approach 
the Maxwellian limit for both Kappa variants. Furthermore, we were able to compare our results to 
those found earlier by \cite{Du-2013}, who used the modified Kappa approach and whose results 
for the thermoelectric coefficient $\alpha$ and the thermal conductivity $\lambda$ differ from our 
results due to a missing term in the derivation of the expression for the perturbed distribution function 
term (see Appendix~\ref{app:correction}).

In the present analysis, we have not incorporated data from measurements or specific values for the temperature 
or associated quantities, but only evaluated normalized transport coefficients. For  normalization, we used the Maxwellian limit that approaches the core of the Kappa distribution, and thus quantifies the contribution of the suprathermal electrons to the transport coefficients. Nonetheless, we have to point out that when using 
observational data, one needs to be cautious whether the measured temperature is after a Kappa fitting as the second-order velocity moment, 
or whether it is the lower one obtained after a Maxwellian fitting with the core of the observed distribution.
In future work we plan to extend these evaluations of the transport coefficients for plasmas described by the  
regularized Kappa distribution (RKD; \cite{Scherer-etal-2017}), which has been introduced to 
formulate a contradiction-free macroscopic description of nonthermal populations observed in the solar wind. By contrast to a standard Kappa, the RKD is defined 
for all values of $\kappa > 0$  
and so are the macroscopic parameters, drift velocity, temperature, heat flux, etc., avoiding limitations corresponding to singularities \citep{Lazar-etal-2020}. 

\begin{acknowledgements}
The authors acknowledge support from the Katholieke Universiteit Leuven and Ruhr-University Bochum. These   results were  obtained  in  the  framework  of  the  projects  SCHL540201/35-1 (DFG-German Research Foundation), C14/19/089  (C1 project Internal Funds KU Leuven), G.0D07.19N  (FWO-Vlaanderen),
SIDC Data Exploitation (ESA Prodex-12), and Belspo projects BR/165/A2/CCSOM and B2/191/P1/SWiM.
\end{acknowledgements}

\bibliographystyle{aa} 
\bibliography{tc_ref}

\begin{thebibliography}{55}
\expandafter\ifx\csname natexlab\endcsname\relax\def\natexlab#1{#1}\fi

\bibitem[{{Alexandrova} {et~al.}(2013){Alexandrova}, {Chen}, {Sorriso-Valvo},
  {Horbury}, \& {Bale}}]{Alexandrova-etal-2013}
{Alexandrova}, O., {Chen}, C.~H.~K., {Sorriso-Valvo}, L., {Horbury}, T.~S., \&
  {Bale}, S.~D. 2013, Space Sci. Rev., 178, 101

\bibitem[{{Bale} {et~al.}(2009){Bale}, {Kasper}, {Howes}, {Quataert}, {Salem},
  \& {Sundkvist}}]{Bale-etal-2009}
{Bale}, S.~D., {Kasper}, J.~C., {Howes}, G.~G., {et~al.} 2009, PRL, 103, 211101

\bibitem[{{Bale} {et~al.}(2013){Bale}, {Pulupa}, {Salem}, {Chen}, \&
  {Quataert}}]{Bale-etal-2013}
{Bale}, S.~D., {Pulupa}, M., {Salem}, C., {Chen}, C.~H.~K., \& {Quataert}, E.
  2013, ApJ Lett., 769, L22

\bibitem[{{Balescu}(1988)}]{Balescu-1988}
{Balescu}, R. 1988, {Transport processes in plasmas - 1. Classical transport
  theory} (Amsterdam: North Holland Publishing Company)

\bibitem[{{Bhatnagar} {et~al.}(1954){Bhatnagar}, {Gross}, \&
  {Krook}}]{Bhatnagar-etal-1954}
{Bhatnagar}, P.~L., {Gross}, E.~P., \& {Krook}, M. 1954, Phys. Rev., 94, 511

\bibitem[{{Bourouaine} {et~al.}(2012){Bourouaine}, {Alexandrova}, {Marsch}, \&
  {Maksimovic}}]{Bourouaine-etal-2012}
{Bourouaine}, S., {Alexandrova}, O., {Marsch}, E., \& {Maksimovic}, M. 2012,
  ApJ, 749, 102

\bibitem[{{Bourouaine} {et~al.}(2011){Bourouaine}, {Marsch}, \&
  {Neubauer}}]{Bourouaine-etal-2011}
{Bourouaine}, S., {Marsch}, E., \& {Neubauer}, F.~M. 2011, A\&A, 536, A39

\bibitem[{{Boyd} \& {Sanderson}(2003)}]{Boyd-Sanderson-2003}
{Boyd}, T.~J.~M. \& {Sanderson}, J.~J. 2003, {The Physics of Plasmas}
  (Cambridge: Cambridge University Press)

\bibitem[{{Braginskii}(1965)}]{Braginskii-1965}
{Braginskii}, S.~I. 1965, Rev. Plasma Phys., 1, 205

\bibitem[{{Collier} {et~al.}(1996){Collier}, {Hamilton}, {Gloeckler},
  {Bochsler}, \& {Sheldon}}]{Collier-etal-1996}
{Collier}, M.~R., {Hamilton}, D.~C., {Gloeckler}, G., {Bochsler}, P., \&
  {Sheldon}, R.~B. 1996, Geophys. Res. Lett., 23, 1191

\bibitem[{{Du}(2013)}]{Du-2013}
{Du}, J. 2013, Phys. Plasmas, 20, 092901

\bibitem[{{Ebne Abbasi} \&
  {Esfandyari-Kalejahi}(2019)}]{Abbasi-Esfandyari-2019}
{Ebne Abbasi}, Z. \& {Esfandyari-Kalejahi}, A. 2019, Phys. Plasmas, 26, 012301

\bibitem[{{Ebne Abbasi} {et~al.}(2017){Ebne Abbasi}, {Esfandyari-Kalejahi}, \&
  {Khaledi}}]{Abbasi-etal-2017}
{Ebne Abbasi}, Z., {Esfandyari-Kalejahi}, A., \& {Khaledi}, P. 2017, Astrophys.
  Space Sci., 362, 103

\bibitem[{{Goedbloed} {et~al.}(2019){Goedbloed}, {Keppens}, \&
  {Poedts}}]{Goedbloed-etal-2019}
{Goedbloed}, H., {Keppens}, R., \& {Poedts}, S. 2019, {Magnetohydrodynamics of
  Laboratory and Astrophysical Plasmas} (Cambridge: Cambridge University Press)

\bibitem[{{Guo} \& {Du}(2019)}]{Guo-Du-2019}
{Guo}, R. \& {Du}, J. 2019, Physica A, 523, 156

\bibitem[{{Gurgiolo} {et~al.}(2016){Gurgiolo}, {Goldstein}, {Vi\~ nas}, \&
  {Fazakerley}}]{Gurgiolo-etal-2016}
{Gurgiolo}, C., {Goldstein}, M.~L., {Vi\~ nas}, A.~F., \& {Fazakerley}, A.~N.
  2016, Ann. Geophys., 34, 1175

\bibitem[{{Hagelaar} \& {Pitchford}(2005)}]{Hagelaar-Pitchford-2005}
{Hagelaar}, G.~J.~M. \& {Pitchford}, L.~C. 2005, Plasma Sources Sci. Technol.,
  14, 722

\bibitem[{{Helander} \& {Sigmar}(2005)}]{Helander-Sigmar-2005}
{Helander}, P. \& {Sigmar}, D.~J. 2005, {Collisional Transport in Magnetized
  Plasmas} (Cambridge: Cambridge University Press)

\bibitem[{{Lazar}(2017)}]{Lazar-2017}
{Lazar}, M. 2017, Phys. Plasmas, 24, 034501

\bibitem[{{Lazar} \& {Fichtner}(2021)}]{Lazar-Fichtner-2021}
{Lazar}, M. \& {Fichtner}, H. 2021, in Kappa Distributions: From Observational
  Evidences via Controversial Predictions to a Consistent Theory of
  Nonequilibrium Plasmas, ed. M.~{Lazar} \& H.~{Fichtner} (Cham: Springer
  Nature)

\bibitem[{{Lazar} {et~al.}(2016){Lazar}, {Fichtner}, \&
  {Yoon}}]{Lazar-etal-2016}
{Lazar}, M., {Fichtner}, H., \& {Yoon}, P.~H. 2016, A\&A, 589, A39

\bibitem[{{Lazar} {et~al.}(2018){Lazar}, {Kim}, {L{\'o}pez}, {Yoon},
  {Schlickeiser}, \& {Poedts}}]{Lazar-etal-2018}
{Lazar}, M., {Kim}, S., {L{\'o}pez}, R.~A., {et~al.} 2018, ApJ Lett., 868, L25

\bibitem[{{Lazar} {et~al.}(2015){Lazar}, {Poedts}, \&
  {Fichtner}}]{Lazar-etal-2015}
{Lazar}, M., {Poedts}, S., \& {Fichtner}, H. 2015, A\&A, 582, A124

\bibitem[{{Lazar} {et~al.}(2013){Lazar}, {Poedts}, \&
  {Michno}}]{Lazar-etal-2013}
{Lazar}, M., {Poedts}, S., \& {Michno}, M.~J. 2013, A\&A, 554, A64

\bibitem[{{Lazar} {et~al.}(2011){Lazar}, {Poedts}, \&
  {Schlickeiser}}]{Lazar-etal-2011}
{Lazar}, M., {Poedts}, S., \& {Schlickeiser}, R. 2011, A\&A, 534, A116

\bibitem[{{Lazar} {et~al.}(2020){Lazar}, {Scherer}, {Fichtner}, \&
  {Pierrard}}]{Lazar-etal-2020}
{Lazar}, M., {Scherer}, K., {Fichtner}, H., \& {Pierrard}, V. 2020, A\&A, 634,
  A20

\bibitem[{{Lazar} {et~al.}(2012){Lazar}, {Schlickeiser}, \&
  {Poedts}}]{Lazar-etal-2012a}
{Lazar}, M., {Schlickeiser}, R., \& {Poedts}, S. 2012, in Exploring the Solar
  Wind, ed. M.~{Lazar} (Rijeka: Intechopen Publishing), 241

\bibitem[{{Maksimovic} {et~al.}(1997){Maksimovic}, {Pierrard}, \&
  {Lemaire}}]{Maksimovic-etal-1997}
{Maksimovic}, M., {Pierrard}, V., \& {Lemaire}, J.~F. 1997, A\&A, 324, 725

\bibitem[{{Maksimovic} {et~al.}(2005){Maksimovic}, {Zouganelis}, {Chaufray},
  {Issautier}, {Scime}, {Littleton}, {Marsch}, {McComas}, {Salem}, {Lin}, \&
  {Elliott}}]{Maksimovic-etal-2005}
{Maksimovic}, M., {Zouganelis}, I., {Chaufray}, J.~Y., {et~al.} 2005, J.
  Geophys. Res., 110, A09104

\bibitem[{{Marsch}(2006)}]{Marsch-2006}
{Marsch}, E. 2006, Liv. Rev. Solar Phys., 3, 1

\bibitem[{{Mason} \& {Gloeckler}(2012)}]{Mason-Gloeckler-2012}
{Mason}, G.~M. \& {Gloeckler}, G. 2012, Space Sci. Rev., 172, 241

\bibitem[{{Matteini} {et~al.}(2007){Matteini}, {Landi}, {Hellinger},
  {Pantellini}, {Maksimovic}, {Velli}, {Goldstein}, \&
  {Marsch}}]{Matteini-etal-2007}
{Matteini}, L., {Landi}, S., {Hellinger}, P., {et~al.} 2007, Geophys. Res.
  Lett., 34, L20105

\bibitem[{{Ogilvie} {et~al.}(1993){Ogilvie}, {Geiss}, {Gloeckler},
  {Berdichevsky}, \& {Wilken}}]{Ogilvie-etal-1993}
{Ogilvie}, K.~W., {Geiss}, J., {Gloeckler}, G., {Berdichevsky}, D., \&
  {Wilken}, B. 1993, J. Geophys. Res., 98, 3605

\bibitem[{{Olbert}(1968)}]{Olbert-1968}
{Olbert}, S. 1968, Phys. Magnetos., 10, 641

\bibitem[{{Pierrard} \& {Lazar}(2010)}]{Pierrard-Lazar-2010}
{Pierrard}, V. \& {Lazar}, M. 2010, Sol. Phys., 267, 153

\bibitem[{{Pierrard} {et~al.}(2011){Pierrard}, {Lazar}, \&
  {Schlickeiser}}]{Pierrard-etal-2011}
{Pierrard}, V., {Lazar}, M., \& {Schlickeiser}, R. 2011, Sol. Phys., 269, 421

\bibitem[{{Rat} {et~al.}(2001){Rat}, {Andr\'e}, {Aubreton}, {Elchinger},
  {Fauchais}, \& {Lefort}}]{Rat-etal-2001}
{Rat}, V., {Andr\'e}, P., {Aubreton}, J., {et~al.} 2001, Phys. Rev. E, 64,
  026409

\bibitem[{{Roberts} \& {Miller}(1998)}]{Roberts-Miller-1998}
{Roberts}, D.~A. \& {Miller}, J.~A. 1998, Geophys. Res. Lett., 25, 607

\bibitem[{{Salem} {et~al.}(2003){Salem}, {Hubert}, {Lacombe}, {Bale},
  {Mangeney}, {Larson}, \& {Lin}}]{Salem-etal-2003}
{Salem}, C., {Hubert}, D., {Lacombe}, C., {et~al.} 2003, ApJ, 585, 1147

\bibitem[{{Scherer} {et~al.}(2017){Scherer}, {Fichtner}, \&
  {Lazar}}]{Scherer-etal-2017}
{Scherer}, K., {Fichtner}, H., \& {Lazar}, M. 2017, Europhys. Lett., 120, 50002

\bibitem[{{Scherer} {et~al.}(2020){Scherer}, {Husidic}, {Lazar}, \&
  {Fichtner}}]{Scherer-etal-2020}
{Scherer}, K., {Husidic}, E., {Lazar}, M., \& {Fichtner}, H. 2020, MNRAS, 497,
  1738

\bibitem[{{Shaaban} {et~al.}(2019){Shaaban}, {Lazar}, {L{\'o}pez}, {Fichtner},
  \& {Poedts}}]{Shaaban-etal-2019}
{Shaaban}, S.~M., {Lazar}, M., {L{\'o}pez}, R.~A., {Fichtner}, H., \& {Poedts},
  S. 2019, MNRAS, 483, 5642

\bibitem[{{Shaaban} {et~al.}(2020){Shaaban}, {Lazar}, {L{\'o}pez}, \&
  {Poedts}}]{Shaaban-etal-2020}
{Shaaban}, S.~M., {Lazar}, M., {L{\'o}pez}, R.~A., \& {Poedts}, S. 2020, ApJ,
  899, 20

\bibitem[{{Spatschek}(1990)}]{Spatschek-1990}
{Spatschek}, K.~H. 1990, {Theoretische Plasmaphysik - Eine Einf\"{u}hrung}
  (Stuttgart: B.~G. Teubner Verlag)

\bibitem[{{\v{S}tver\'{a}k} {et~al.}(2008){\v{S}tver\'{a}k},
  {Tr\'{a}vni\v{c}ek}, {Maksimovic}, {Marsch}, {Fazakerley}, \&
  {Scime}}]{Stverak-etal-2008}
{\v{S}tver\'{a}k}, {\v{S}}., {Tr\'{a}vni\v{c}ek}, P., {Maksimovic}, M.,
  {et~al.} 2008, J. Geophys. Res., 113, 103

\bibitem[{{Vasyli\={u}nas}(1968)}]{Vasyliunas-1968}
{Vasyli\={u}nas}, V.~M. 1968, J. Geophys. Res.: Space Phys., 73, 2839

\bibitem[{{Vi\~nas} {et~al.}(2017){Vi\~nas}, {Gaelzer}, {Mace}, {Moya}, \&
  {Araneda}}]{Vinas-etal-2017}
{Vi\~nas}, A.~F., {Gaelzer}, R., {Mace}, R., {Moya}, P.~S., \& {Araneda}, J.~A.
  2017, in Kappa Distributions, ed. G.~{Livadiotis} (Amsterdam: Elsevier)

\bibitem[{{Vi\~nas} {et~al.}(2015){Vi\~nas}, {Moya}, {Navarro}, {Valdivia},
  {Araneda}, \& {Mu\~noz}}]{Vinas-etal-2015}
{Vi\~nas}, A.~F., {Moya}, P.~S., {Navarro}, R.~E., {et~al.} 2015, J. Geophys.
  Res.: Space Phys., 120, 3307

\bibitem[{{Vi\~nas} {et~al.}(2000){Vi\~nas}, {Wong}, \&
  {Klimas}}]{Vinas-etal-2000}
{Vi\~nas}, A.~F., {Wong}, H.~K., \& {Klimas}, A.~J. 2000, ApJ, 528, 509

\bibitem[{{Vocks} {et~al.}(2008){Vocks}, {Mann}, \&
  {Rausche}}]{Vocks-etal-2008}
{Vocks}, C., {Mann}, G., \& {Rausche}, G. 2008, A\&A, 480, 527

\bibitem[{{Wang} \& {Du}(2017)}]{Wang-Du-2017}
{Wang}, L. \& {Du}, J. 2017, Phys. Plasmas, 24, 102305

\bibitem[{{Wilson} {et~al.}(2019){Wilson}, {Chen}, {Wang}, {Schwartz},
  {Turner}, {Stevens}, {Kasper}, {Osmane}, {Caprioli}, {Bale}, {Pulupa},
  {Salem}, \& {Goodrich}}]{Wilson-etal-2019}
{Wilson}, L.~B., {Chen}, L.~J., {Wang}, S., {et~al.} 2019, ApJS, 243, 8

\bibitem[{{Wilson} {et~al.}(2013){Wilson}, {Koval}, {Szabo}, {Breneman},
  {Cattell}, {Goetz}, {Kellogg}, {Kersten}, {Kasper}, {Maruca}, \&
  {Pulupa}}]{Wilson-etal-2013}
{Wilson}, L.~B., {Koval}, A., {Szabo}, A., {et~al.} 2013, J. Geophys. Res.,
  118, 5

\bibitem[{{Yoon} {et~al.}(2006){Yoon}, {Rhee}, \& {Ryu}}]{Yoon-etal-2006}
{Yoon}, H.~P., {Rhee}, T., \& {Ryu}, C.~M. 2006, J. Geophys. Res., 111, A09106

\bibitem[{{Zouganelis} {et~al.}(2005){Zouganelis}, {Meyer-Vernet}, {Landl},
  {Maksimovic}, \& {Pantellini}}]{Zouganelis-etal-2005}
{Zouganelis}, I., {Meyer-Vernet}, N., {Landl}, S., {Maksimovic}, M., \&
  {Pantellini}, F. 2005, ApJ, 626, L117

\end{thebibliography}

\begin{appendix} 
\section{Comments on the results in Du (2013)} \label{app:correction}

In the derivation of the transport coefficients, 
we noticed a missing term in Eq.~(10) in \cite{Du-2013} (readers are encouraged to compare this to Eq.~\eqref{eq:f_1} from the present paper).
This term appears in both approaches and it is not clear why this term is left out in Du (2013), that is, whether it was overlooked in the derivation or neglected for simplification.
While this has no effect on the
electrical conductivity ($\sigma$) calculated in \cite{Du-2013}, we obtain different results for the thermoelectric 
coefficient ($\alpha$) and the thermal conductivity ($\lambda$) when considering the missing term. 

\begin{figure}[t]
   \includegraphics[width=\hsize]{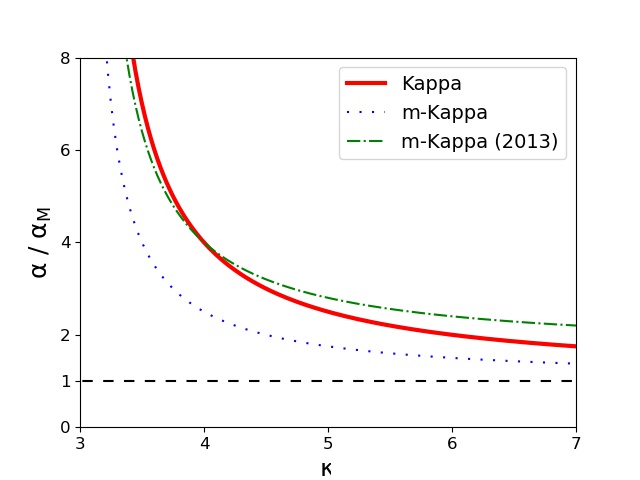}
   \caption{Results for the thermoelectric coefficient $\alpha$ from Fig.~\ref{fig:alpha} and those from \cite{Du-2013} for the modified Kappa approach (m-Kappa 2013, green dash-dotted line).}        \label{fig:app_alpha}%
    \end{figure}

\begin{figure}[t]
   \includegraphics[width=\hsize]{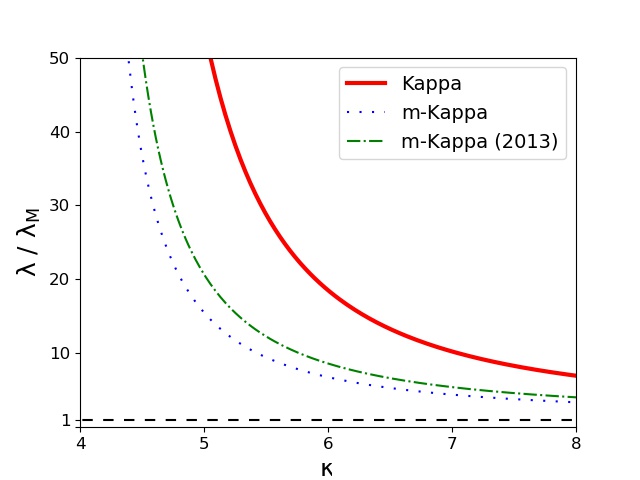}
   \caption{Results for the thermal conductivity coefficient $\lambda$ from Fig.~\ref{fig:lambda} and those from \cite{Du-2013} for the modified Kappa approach (m-Kappa 2013, green dash-dotted line).} \label{fig:app_lambda}%
    \end{figure}

In \cite{Du-2013}, derivations are based on the modified Kappa approach, and the results are 
\begin{equation}
    \alpha = -\frac{\kappa - 3/2}{\kappa - 3}\, 4\, \frac{k_\mathrm{B}}{e}\,
,\end{equation}
which is slightly different from our Eq.~\eqref{eq:alpha} above for $\eta = \kappa-3/2$, and 
\begin{equation}
    \lambda = \frac{\kappa + 1}{\kappa -3}\,
    \frac{\Gamma(\kappa - 4)}{\Gamma\left(\kappa - 1/2\right)}\, \left(\kappa - \frac{3}{2} \right)^{7/2}
    \frac{16\,\sqrt{2}\,k_\mathrm{B}\,\left(k_\mathrm{B}\,T \right)^{5/2}}
    {m^{1/2}\,\pi^{3/2}\,z\,e^4\,L_\mathrm{ei}}
,\end{equation}
which is also different from our Eq.~\eqref{eq:lambda} above for $\eta = \kappa-3/2$.
These two expressions are plotted (green dash-dotted lines) in Figs~\ref{fig:app_alpha} and 
\ref{fig:app_lambda}, respectively, by contrast to our results. The results in \cite{Du-2013} differ from the values obtained in the present work for the modified Kappa approach (blue-dotted lines), and also from those obtained for a standard Kappa approach (red lines).
%

\section{Integrals} \label{app:Integrals}
The integrals occurring in Sec.~\ref{sec:transport_coefficients} can be calculated by using
substitution and integration by parts. \cite{Scherer-etal-2020} derived a general form to calculate
integrals of type
\begin{equation}\label{eq:appendix_general_integral}
    \Big \langle v^n\, (1 + A_\kappa\,v^2)^\zeta \Big \rangle = 
    4\,\pi\,N_\kappa \int\limits_{0}^{\infty} \mathrm{d}v\, v^{\,n+2}\,(1 + A_\kappa v^2)^{\zeta\, - \,(\kappa + 1)}
,\end{equation}
using spherical coordinates and having already calculated the integrals over $\vartheta$ and $\phi$. For the integrals, we then obtain the following expressions:
\begin{align}
    \Big \langle v^5\, (1 + A_\kappa\,v^2)^{-1} \Big \rangle &= 12\,\pi \frac{N_\kappa}{A_\kappa^4} \frac{\Gamma(\kappa - 2)}{\Gamma(\kappa + 2)}\,, \label{eq:app_b_1} \\
    \Big \langle v^7\, (1 + A_\kappa\,v^2)^{-1} \Big \rangle &=  48\,\pi \frac{N_\kappa}{A_\kappa^5} \frac{\Gamma(\kappa - 3)}{\Gamma(\kappa + 2)}\,, \label{eq_app_b_2} \\
    \Big \langle v^9\, (1 + A_\kappa\,v^2)^{-1} \Big \rangle &= 240\,\pi \frac{N_\kappa}{A_\kappa^6} \frac{\Gamma(\kappa - 4)}{\Gamma(\kappa + 2)}\,, \label{eq:app_b_3} \\
    \Big \langle v^5 \Big \rangle &= 12\,\pi \frac{N_\kappa}{A_\kappa^4} 
    \frac{\Gamma(\kappa - 3)}{\Gamma(\kappa + 1)}\,, \label{eq:app_b_4} \\
    \Big \langle v^7 \Big \rangle &= 48\,\pi \frac{N_\kappa}{A_\kappa^5} 
    \frac{\Gamma(\kappa - 4)}{\Gamma(\kappa + 1)}\,. \label{eq:app_b_5}
\end{align}
%

\section{Tabulated transport coefficients} \label{app:values}

In this appendix we present a series of selected values of the normalized coefficients represented graphically in Figs. 1 -- 5 in order to quantify
the differences introduced by the suprathermals in the standard Kappa approach (subscript K) by comparison to the Maxwellian (subscript M), and also the
underestimations from using a modified Kappa approach (subscript m-K).

The numbers in the tables clearly indicate the reinforcing effect on all transport coefficients by the presence of suprathermal particles represented by the original Kappa distribution $f_\mathrm{K}$. For the electric conductivity $\sigma$ (mobility coefficient $\mu$, Tabs.~\ref{tab:c1_sigma_mu} and~\ref{tab:c5_mu}) and the thermoelectric coefficient $\alpha$ (Tab.~\ref{tab:c2_alpha}), we find values 7 times the Maxwellian at $\kappa = 2.5$ and $\kappa = 3.5$, respectively. The contribution of suprathermal particles to the thermal conductivity $\lambda$ (Tab.~\ref{tab:c3_lambda}) and the diffusion coefficient $D$ (Tab.~\ref{tab:c4_diffusion}) is even greater, and they manifest in values as large as about 18 to 150 times the Maxwellian for $\lambda$ (at $\kappa = 4.5$ to $6$) and about 10 to 20 times the Maxwellian for $D$ (at $\kappa = 3.5$ to $4$).

By comparing the original Kappa approach to the modified one ($f_\mathrm{m-K}$), we identify an underestimation of all transport coefficients by the latter. Although the values of $\lambda$ based on $f_\mathrm{m-K}$ are still relatively large (4 to 36 times the Maxwellian-based results) at values $\kappa < 8$, they are 2 to 4 times smaller than the results obtained with $f_\mathrm{K}$.
The results of the remaining transport coefficients reveal an even weaker contribution by suprathermal particles represented by $f_\mathrm{m-K}$ in comparison to the original Kappa approach, for example, for $\sigma$ ($\mu$) less than twice the Maxwellian at values for Kappa as small as $\kappa = 2.5$, and 4 times smaller than the $f_\mathrm{K}$-based value. For $\alpha$ and $D$, the results obtained with $f_\mathrm{m-K}$ are less than twice the Maxwellian for $\kappa \gtrsim 5$, and again 2 to 4 times smaller than the numbers resulting from $f_\mathrm{K}$.

\begin{table}[h!]
\caption{Numerical values of $\sigma$.} \label{tab:c1_sigma_mu}
\centering
\begin{tabular}{cccccc}
\hline
\hline
 $\kappa$ & 2.5 & 3 & 4 & 5 & 6 \\
\hline 
$\sigma_{\rm K}/\sigma_{\rm M}$ & 7.01 & 3.91 & 2.41 & 1.92 & 1.69 \\
$\sigma_{\rm m-K}/ \sigma_{\rm M}$  & 1.77 & 1.38 & 1.19 & 1.13 & 1.09 \\
\hline
\end{tabular} 
\tablefoot{Selected values of the normalized electric conductivity $\sigma$ plotted in Fig.~\ref{fig:sigma}.}
\end{table}

\begin{table}[h!]
\caption{Numerical values of $\alpha$.} \label{tab:c2_alpha}
\centering
\begin{tabular}{cccccc}
\hline
\hline
 $\kappa$ & 3.5 & 4 & 5 & 6 & 7 \\
\hline 
$\alpha_{\rm K}/\alpha_{\rm M}$ & 7 & 4 & 2.5 & 2 & 1.75 \\
$\alpha_{\rm m-K}/ \alpha_{\rm M}$  & 4 & 2.5 & 1.75 & 1.5 & 1.38 \\
\hline
\end{tabular} 
\tablefoot{Selected values of the normalized thermoelectric coefficient $\alpha$ plotted in Fig.~\ref{fig:alpha}.}
\end{table}

\begin{table}[h!]
\caption{Numerical values of $\lambda$.} \label{tab:c3_lambda}
\centering
\begin{tabular}{cccccc}
\hline
\hline
 $\kappa$ & 4.5 & 5 & 6 & 7 & 8 \\
\hline 
$\lambda_{\rm K}/\lambda_{\rm M}$ & 152.88 & 54.07 & 18.53 & 10.24 & 6.97 \\
$\lambda_{\rm m-K}/ \lambda_{\rm M}$  & 36.84 & 15.52 & 6.77 & 4.40 & 3.37 \\
\hline
\end{tabular} 
\tablefoot{Selected values of the normalized  thermal conductivity $\lambda$ plotted in Fig.~\ref{fig:lambda}.}
\end{table}

\begin{table}[h!]
\caption{Numerical values of $D$.} \label{tab:c4_diffusion}
\centering
\begin{tabular}{cccccc}
\hline
\hline
 $\kappa$ & 3.5 & 4 & 5 & 6 & 7 \\
\hline 
$D_{\rm K}/D_{\rm M}$ & 20.31 & 9.63 & 4.81 & 3.37 & 2.70 \\
$D_{\rm m-K}/ D_{\rm M}$  & 5.01 & 2.97 & 1.97 & 1.64 & 1.48 \\
\hline
\end{tabular} 
\tablefoot{Selected values of the normalized diffusion coefficient $D$ plotted in Fig.~\ref{fig:diffusion}.}
\end{table}

\begin{table}[h!]
\caption{Numerical values of $\mu$.} \label{tab:c5_mu}
\centering
\begin{tabular}{cccccc}
\hline
\hline
 $\kappa$ & 2.5 & 3 & 4 & 5 & 6 \\
\hline 
$\mu_{\rm K}/\mu_{\rm M}$ & 7.01 & 3.91 & 2.41 & 1.92 & 1.69 \\
$\mu_{\rm m-K}/ \mu_{\rm M}$  & 1.77 & 1.38 & 1.19 & 1.13 & 1.09 \\
\hline
\end{tabular} 
\tablefoot{Selected values of the normalized mobility coefficient $\mu$ plotted in Fig.~\ref{fig:mobility}. Due to the normalization to the Maxwellian limit, these values are the same as those obtained for the electric conductivity $\sigma$ in Table \ref{tab:c1_sigma_mu}. }
\end{table}

\end{appendix}
\end{document}